\documentclass[%
 reprint,
 superscriptaddress,
 amsmath,amssymb,
 aps,
 pra,
floatfix,
nofootinbib
]{revtex4-2}

\usepackage[T1]{fontenc}
\usepackage{mathptmx}
\usepackage{graphicx}
\usepackage{dcolumn}
\usepackage{physics}
\usepackage{bm}
\usepackage{hyperref}
\hypersetup{
    colorlinks=true,
    linkcolor=blue,
    filecolor=blue,      
    urlcolor=blue,
    citecolor=blue
    }

\begin{document}

\title{Probing early phase coarsening in a rapidly quenched Bose gas using off-resonant matter-wave interferometry}

\author{Tenzin Rabga}\email{trabga@snu.ac.kr}
\affiliation{Department of Physics and Astronomy, Seoul National University, Seoul 08826, Korea}
\author{Yangheon Lee}
\affiliation{Department of Physics and Astronomy, Seoul National University, Seoul 08826, Korea}
\author{Y. Shin}\email{yishin@snu.ac.kr}
\affiliation{Department of Physics and Astronomy, Seoul National University, Seoul 08826, Korea}
\affiliation{Institute of Applied Physics, Seoul National University, Seoul 08826, Korea}
 
\date{\today}

\begin{abstract}
We experimentally investigate the evolution of spatial phase correlations in a rapidly quenched inhomogeneous Bose gas of rubidium using off-resonant matter-wave interferometry. We measure the phase coherence length $\ell$ of the sample and directly probe its increase during the early stage of condensate growth before vortices are formed. Once the vortices are formed stably in the quenched condensate, the measured value of $\ell$ is shown to be linearly proportional to the mean distance between the vortex. These results confirm the presence of phase coarsening prior to vortex formation, which is crucial for a quantitative understanding of the resultant defect density in samples undergoing critical phase transitions.
\end{abstract}

\maketitle

\section{Introduction} 

Spontaneous symmetry-breaking phase transitions have been widely studied both theoretically and experimentally. Understanding the out-of-equilibrium dynamics of a system in the vicinity of the critical point and the resultant emergence of an ordered phase is a key question in quantum many-body physics. It is also crucial for quantum state preparation in applications ranging from quantum information processing to quantum sensing and metrology, where deterministic control over internal and motional states for a large parameter space is highly desirable.

The Kibble-Zurek mechanism (KZM) provides a universal framework for understanding the non-equilibrium dynamics of a system undergoing a second-order phase transition~\cite{Kibble_1976,Zurek1985,ZUREK1996177}. For a thermal transition, as the temperature $T$ approaches the critical temperature $T_{c}$, the dynamics of the system is characterized by the divergence of the correlation length $\xi \propto \abs{T-T_{c}}^{-\nu}$ and the relaxation time $\tau \propto \xi^{z}$, where $\nu$ and $z$ are the static and dynamical critical exponents, respectively, that are independent of the microscopic details of the system, and are solely determined by its universality class. At a certain timescale $\hat{t}$ before reaching the critical point, the system dynamics critically slows down and the growth of the correlation length ceases. This frozen-out correlation length $\hat{\xi}$ sets the characteristic size of the phase domains over which the phase of the broken-symmetry state is chosen independently. The KZM predicts a power-law scaling of $\hat{\xi}$ with the quench time $t_{q}$ over which the system is driven across the critical point: $\hat{\xi} \propto t_{q}^{\nu/(1+\nu z)}$. Consequently, the merging of these domains leads to the formation of topological defects in the system whose number density $n_{v}$ also displays a power-law dependence on the quench time~\cite{del_Campo_2013,Dziarmaga10}.

KZM predictions have been extensively tested in ultracold atomic gas experiments~\cite{Lamporesi2013,Chomaz15,Goo21,rabga23,Navon15,Ko2019,Donadello16,Lee23}. In a homogeneous Bose gas of rubidium, using matter-wave interferometry~\cite{stenger99, hagley99, hugbart05}, the underlying correlation length was measured for various quench times and shown to have the predicted power law scaling~\cite{Navon15}. This strongly validated the freeze-out hypothesis, a core concept in KZM. In experiments using large-area samples of highly oblate geometry, the number of quantum vortices generated was directly counted, revealing the characteristic power-law scaling of the defect number with the quench time in the relatively slow quench regime~\cite{Lamporesi2013,Goo21,rabga23,Ko2019,Donadello16}. Recently, in quasi-two-dimensional (2D) homogeneous Fermi gases, the KZ scaling exponent for the defect number was measured, free of the inhomogeneous density effects, and was shown to be consistent with theoretical predictions~\cite{Lee23}.

Nevertheless, these vortex counting experiments reveal an intriguing departure from the KZM; a distinctly clear saturation of the vortex numbers was observed in the fast quench regime, which was attributed to the early-time phase coarsening the initial turbulent condensate experiences~\cite{chesler15}. Furthermore, regardless of the details of sample preparation, such as trapping conditions and atomic densities, the observed saturation defect densities, $n_{v,\text{sat}}$, and the threshold saturation quench times, $t_{q,\text{sat}}$, were found to be universal when normalized with respect to the characteristic length and time of the system, respectively, i.e., $n_{v,\text{sat}}\xi_{v}^{2} \approx 2\times10^{-4}$ and $t_{q,\text{sat}}/\tau_{\text{el}} \approx 3\times 10^{3}$, where $\xi_{v}$ is the characteristic vortex core size and $\tau_{\text{el}}$ is the elastic collision time~\cite{rabga23}. This implies the presence of a different type of universality in the rapid quench regime ($t_q<t_{q,\text{sat}}$), involving coarsening of the phase domains during the early stages of condensate growth and subsequent vortex formation~\cite{chesler15,Gooetal22}. It is noteworthy that the relation $t_{q,\text{sat}}/\tau_{\text{el}} \approx 3\times 10^{3}$ suggests that the quench experiments in \cite{Navon15} were performed in a deep saturation regime since $\tau_{\text{el}} \approx 30$ ms implies an estimated $t_{q,\text{sat}} \approx 100\,\text{s}$. As a result, the defect numbers, if counted, would not have shown any power-law scaling, despite the observed scaling of the correlation length with the quench time. In other words, the early-time coarsening significantly constrains the vortex density to a saturation value in rapidly quenched samples, despite their different initial correlation lengths. In fact, in \cite{Navon15}, the coherence length of the quenched sample was observed to increase after the freeze-out period, hinting at an early-time coarsening before defect formation.

In this paper, we experimentally investigate the evolution of spatial phase correlations in a rapidly quenched Bose gas of rubidium and study its implications for the resultant vortex formation. Using homodyne matter-wave interferometry, we measure the phase coherence lengths $\ell$ of quenched samples and directly probe their increase during the early stage of condensate growth before vortices are formed. Furthermore, we find that once the vortices are stably created in the quenched condensate, the measured value of $\ell$ is related to the mean intervortex distance $d_v$ by $\ell\approx 0.3 d_v$. These results confirm the presence of spatial phase coarsening prior to vortex formation, and validate the early coarsening effect inferred from vortex counting experiments and the observed saturation of defect numbers~\cite{Gooetal22,rabga23}. We emphasize that the early-time coarsening is a process beyond the conventional KZM and is significant for a quantitative understanding of the resultant defect density in the critical phase transition dynamics. 

The rest of the paper is organized as follows. In Sec.~II, we describe our experimental setup and the matter-wave interferometry technique to determine the phase coherence length of the sample~\cite{Navon15,hagley99}. Here, we emphasize the use of off-resonant Bragg scattering, which, in our system, helps circumvent the need to precisely determine the two-photon resonance. In Sec.~III, we present the experimental results and compare the measured coherence lengths to the vortex densities. Crucially, we present and discuss the direct observation of phase coarsening in rapidly quenched samples and its implications for the vortex formation process. Lastly, Sec.~IV provides some concluding remarks.

\section{Experimental Setup}

\begin{figure}[htb]
    \centering
    \includegraphics[width=3.4in]{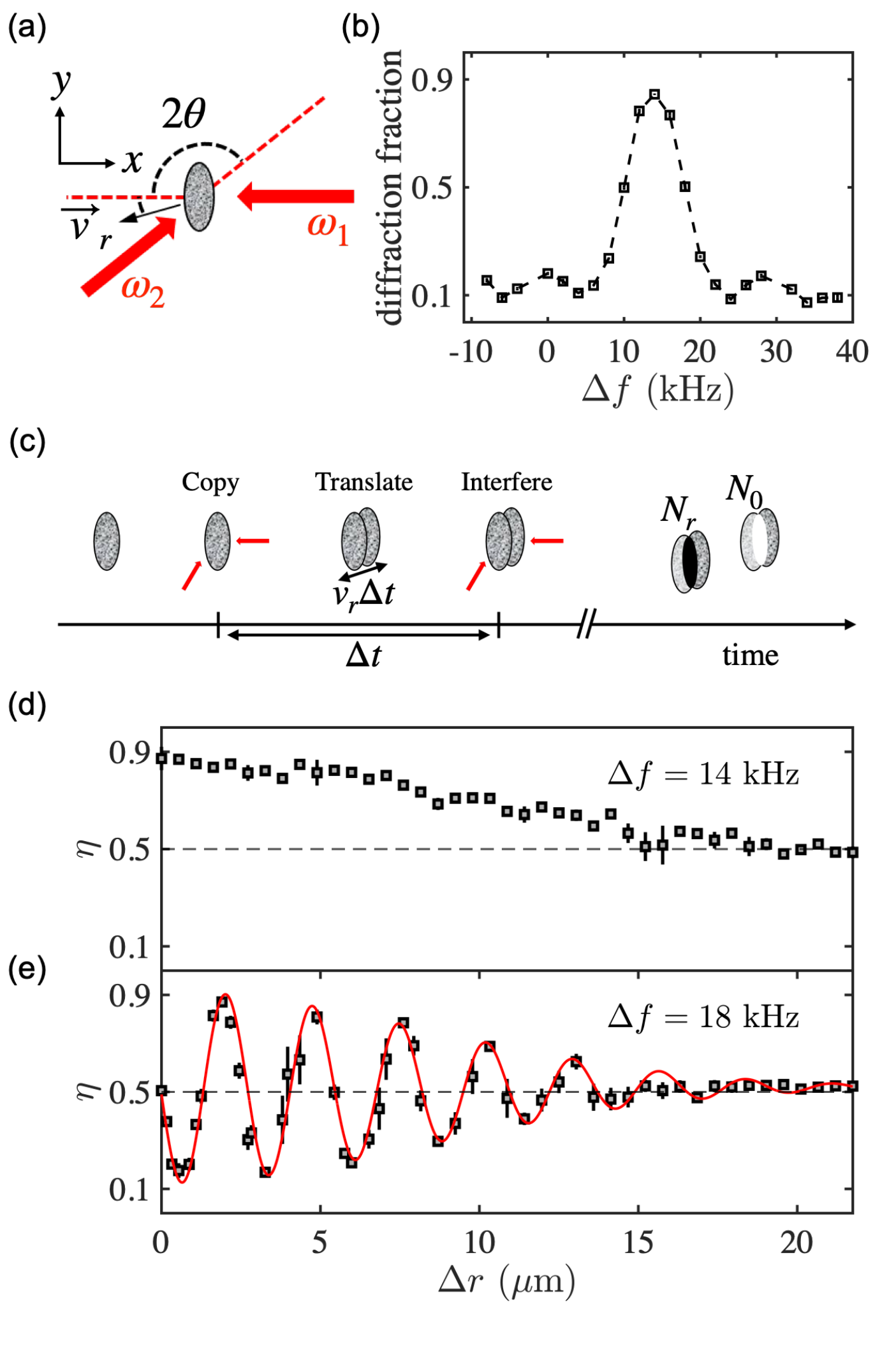}%
    \caption{Matter-wave interferometry using Bragg scattering. (a) Schematic of the Bragg scattering setup. The atoms are irradiated with two beams with frequencies $\omega_{1}$ and $\omega_{2}$ at an angle $2\theta = 3\pi/4$, diffracting a portion of the atoms into a motional state with a recoil velocity $v_{r}=10.9$~mm/s. (b) Diffraction fraction as a function of $\Delta f = (\omega_{1}-\omega_{2})/2\pi$ with resonance observed at $\Delta f \approx 14$ kHz. (c) Matter-wave interferometry sequence. A sample of $N$ atoms is illuminated with a pulse of Bragg beams, transferring $\approx 50\%$ of the atoms to the diffracted state. The two copies are allowed to separate for a time $\Delta t$ before another Bragg pulse is applied. The second beam pulse causes interference of the two copies of the sample in their region of overlap. After an additional time of flight, the atoms in the diffracted copy ($N_{r}$) and the remaining copy ($N_{0}$) are counted to determine the fraction of out-coupled atoms $\eta = N_{r}/N$, where $N = N_{0}+N_{r}$. Interference signal $\eta$ as a function of separation $\Delta r = v_{r}\Delta t$ for (d) $\Delta f=14$~kHz and (e) 18~kHz. The solid line in (e) is the fit to the data using Eq.~(\ref{eq:fitfun}).}
    \label{fig:Braggsetup}
\end{figure}

\subsection{Sample preparation}

We begin with a thermal sample of $^{87}$Rb atoms trapped in an optical dipole trap (ODT). The sample is cooled below the critical temperature $T_{c}$ for Bose-Einstein condensation by linearly lowering the depth of the ODT from an initial depth of $U_{i} = 1.15U_{c}$ to a final depth of $U_{f} = 0.36U_{c}$ in a variable quench time $t_{q}$, where $U_{c}$ is the critical ODT depth corresponding to $T_c$. The quench protocol and the range of $t_{q}$ values probed here is similar to what is shown in ~\cite{Goo21}, where the sample temperature linearly follows the trap depth. Since the underlying elastic collision times are also similar, we assume a linear relationship between the trap depth and temperature in the samples presented in this work. At the end of the quench, we hold the sample in the trap for a time $t_{h}$ to facilitate the growth of condensate. The trapping frequencies of the final ODT are $\omega_{x,y,z}/2\pi = (8.2,3.2,85)$ Hz and the typical number of atoms after the quench is about $10^7$. Additional details of sample preparation can be found in ~\cite{Goo21,rabga23}. Depending on the particular aspect of the phase transition process under investigation, sample preparation can differ. For example, to examine the relation between $\ell$ and $n_v$, we vary $t_{q}$ from 0.5 to 4~s with a fixed $t_{h} = 1.25$~s. In addition, to investigate the early coarsening effect, we set $t_{q} = 0.5$ s and study the dependence of the coherence length on the hold time by varying $t_{h}$ from 0.3 to 5~s. 

\subsection{Matter-wave interferometry}

We employ Bragg scattering-based matter-wave interferometry to measure the phase coherence length of the sample~\cite{Navon15, stenger99, hagley99, hugbart05}. Three states are involved in the Bragg scattering process: the ground state $\ket{g} \equiv 5^{2}S_{1/2}\,(F=1)$ and the motional excited state $\ket{e}$ connected by a two-photon transition, and the intermediate state $\ket{i} \equiv 5^{2}P_{3/2}\,(F=0,1,2)$. As shown in Fig.~\ref{fig:Braggsetup}(a), two Bragg beams with frequencies $\omega_{1}$ and $\omega_{2}$ illuminate a sample of rubidium atoms in state $\ket{g}$, red-detuned from the $5^{2}S_{1/2} \rightarrow 5^{2}P_{3/2}$ transition by $\approx 2\pi \times 6.6$~GHz. A fraction of the original cloud is diffracted into the motional excited state $\ket{e}$. It recoils with a velocity $\vec{v}_{r}$ determined by the geometry of the setup. Using 780 nm Bragg beams intersecting at an angle $2\theta = 3\pi/4$, we estimate a recoil velocity of $v_{r} = 10.9$~mm/s. Figure~\ref{fig:Braggsetup}(b) shows the diffraction fraction of the condensate as a function of the frequency difference between the two Bragg beams, $2\pi\Delta f = \omega_{1}-\omega_{2}$. We find the two-photon resonance at $\Delta f \approx 14$ kHz, which is consistent with the estimated value of the recoil velocity $v_r$.

The experimental sequence for homodyne matter-wave interferometry is shown in Fig.~\ref{fig:Braggsetup}(c). After an initial time-of-flight of 3 ms, we apply the first pulse of Bragg beams to the sample, creating a diffracted copy of the condensate. The pulse width ($\tau = 50~\mu$s) and the beam intensities are chosen such that the diffracted copy contains $\approx 50\%$ of the total atoms. This is crucial to ensure optimal contrast in our matter-wave interference data. Following a variable pulse separation time $\Delta t$ -- during which the two copies of the sample separate by $\Delta r = v_{r}\Delta t$ -- we apply the second Bragg pulse, resulting in the interference of the two condensates in their region of overlap. After an additional time-of-flight of about $35$ ms, allowing the two copies to separate sufficiently, we image the atoms and determine the population fraction in the out-coupled atoms, $\eta = N_{r}/N$, where $N_{r}$ is the number of out-coupled atoms and $N$ is the total number of atoms. 

The phase coherence length of the sample is determined from the decay of this interference signal $\eta$ with increasing $\Delta r$~\cite{Navon15}. As a feature of our matter-wave interferometry, we deliberately set the frequency difference between the two Bragg beams to $\Delta f = 18$ kHz, such that there is a two-photon detuning of $\delta/2\pi \approx 4$ kHz. A non-zero $\delta$ ensures oscillations in the interferometry signal. For example, Figs.~\ref{fig:Braggsetup}(d) and \ref{fig:Braggsetup}(e) show the interferometry signals for two different values of $\Delta f$ for quasi-pure Bose-Einstein condensates (BECs) prepared with $t_{q} = 12$ s. Close to the resonance there is almost no oscillation in the data [Fig.~\ref{fig:Braggsetup}(d)], while oscillations become noticeable as we increase $\delta$ [Fig.~\ref{fig:Braggsetup}(e)]. Therefore, to measure the decay of the interference signal, it is important to precisely determine the value of $\delta$. It is pertinent at this point to stress our use of off-resonant Bragg scattering in this experiment. Unless the resonance is well-resolved, any small detuning can result in slow oscillations in the signal that can mimic the effects of an exponential decay, and thus systematically shift our determination of the coherence length. To reliably obtain the $\delta$ value near the resonance, signal over a significantly larger range of $\Delta t$ would need to be taken than what is feasible in our system owing to the density inhomogeneity of the samples. We avoid these complications by operating the interferometry off resonance with $\Delta f= 18$~kHz. We find that over the range of $\Delta t$ values probed in this experiment, we are able to sufficiently resolve oscillations at this off-resonant setting. Moreover, as is shown in Fig.~\ref{fig:corrlen_vs_thold}(h), dipole oscillations of the samples trapped in an ODT induce modulations of the two-photon resonance  which needs to be measured for each individual choice of $t_{q}$ and $t_{h}$. Therefore, while two-photon resonances can be determined precisely in general, certain features of our samples, such as, the density inhomogeneity and the resultant decay length scale of the interferometry signal, and the dipole oscillation of the samples trapped in an ODT, make it simpler to operate with a non-zero two-photon detuning, which can be directly determined from the interferometry data.

\section{Results and Discussion}

\subsection{Determination of coherence length}

In determining the coherence length $\ell$ of the sample from the interference signal $\eta(\Delta r)$, there are additional sources of signal decay that need to be properly characterized. First, provided the finite size of the sample, a geometrical decay factor must be taken into account, especially when the separation of the two copies of the condensate approaches significant fractions of the spatial extent of the sample. Second, as explained below, the signal acquires an intrinsic decay length scale that results from the expansion dynamics of the condensate due to its underlying density inhomogeneity.

These considerations lead to the following function used to fit the interferometry data:

\begin{align}
\begin{split}
    \eta(\Delta r) = & A \exp-\left(\frac{1}{2}\frac{\Delta r^{2}}{\ell_{\text{bg}}^{2}} + \frac{\Delta r}{\ell}\right)\cos\left(\frac{\omega}{v_{r}}\Delta r + \phi\right) + B,
\label{eq:fitfun}
\end{split}
\end{align}
where $A$ is the oscillation amplitude, $\ell_{\text{bg}}$ is the ``background'' decay length, $\ell$ is the quantity of interest, i.e., the phase coherence length, $\omega$ is the oscillation frequency, $\phi$ is a phase offset, and $B$ is the signal offset. The exponential term contains the two decay length scales $\ell_{\text{bg}}$ and $\ell$, where $\ell_\text{bg}$ corresponds to the intrinsic signal decay in the system, independent of $t_{q}$. The cosine term results from a nonzero two-photon detuning $\delta$. A detailed derivation of this fit function can be found in Appendix~\ref{appa}. While $A$, $\ell$, $\omega$, $\phi$, and $B$ are parameters obtained from the fit, $\ell_{\text{bg}}$ and $v_{r}$ are predetermined. 

\begin{figure}[t]
    \centering
    \includegraphics[width=3.3in]{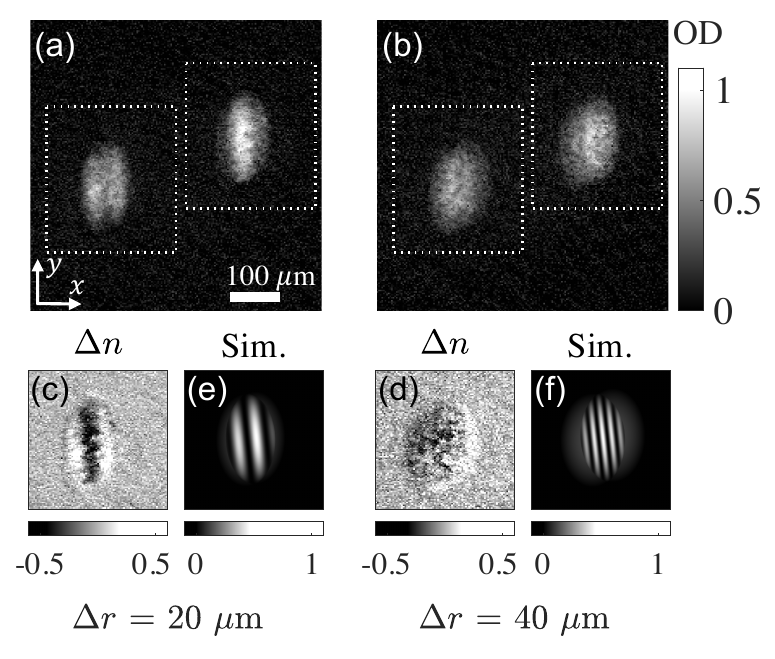}%
    \caption{Appearance of spatial interference pattern in the interfering BECs. Experimental image data of the matter-wave interferometry for sample separations (a) $\Delta r = 20\, \mu$m and (b) $40\,\mu$m.  The initial samples were prepared with $t_q = 10$~s. (c,d) Corresponding density difference distributions $\Delta n(\vec{r})$ of the out-coupled atomic cloud and the remaining atomic cloud. The density difference was calculated as $\Delta n(\vec{r})=n_r(\vec{r}+\vec{r}_D)-n_0(\vec{r})$, where $n_{r(0)}(\vec{r})$ is the density distribution of the out-coupled (remaining) cloud and $\vec{r}_D$ is the displacement of the out-coupled atomic cloud with respect to the remaining one. (e,f) Simulated images of the density distributions of the out-coupled BECs for the two sample separations, respectively (see Appendix~\ref{appb} for details).}
    \label{fig:fringes}
\end{figure}

\begin{figure}[t]
    \centering
    \includegraphics[width=3.3in]{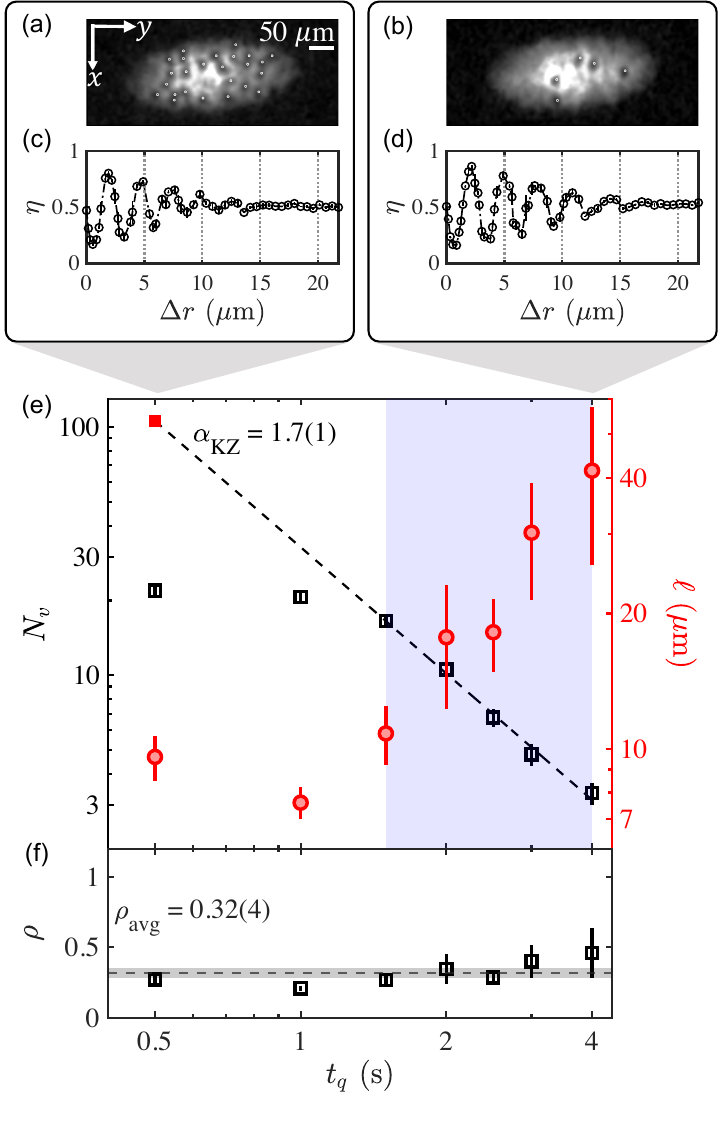}%
    \caption{Vortex number $N_v$ versus coherence length $\ell$. Time-of-flight images of the condensates containing vortices for (a) $t_{q} = 0.5$~s and (b) 4~s. The white dots indicate the positions of the vortices in the samples. (c,d) Corresponding interferometry signals $\eta$ as functions of $\Delta r$. (e) Mean vortex number $N_{v}$ (black squares) and coherence length $\ell$ (red circles) as functions of $t_{q}$. Each $N_v$ data point is the average of 20 measurements, and the error bar indicates the standard error of the mean. The error bar for each $\ell$ is the fitting error obtained from the fit of Eq.~(\ref{eq:fitfun}) to the respective interferometry data. The dashed line is the power law fit to the $N_v$ data in the shaded region with a scaling exponent $\alpha^{}_{\text{KZ}} = 1.7(1)$, and the red square indicates the extrapolated vortex number at $t_{q} = 0.5$~s. (f) Length ratio $\rho = \ell/d_{v}$ for different values of $t_{q}$, where $d_{v} = \sqrt{\pi R_{x}R_{y}/N_{v}}$ is the intervortex separation with $R_{x(y)}$ being the radius of the condensate along $x(y)$ direction. The dashed line indicates the mean value of $\rho$ and the shaded region indicates the standard error of the mean.}
    \label{fig:corrlen_vs_qtime}
\end{figure}

The value of $\ell_{\text{bg}}$ is determined from the interference signal obtained for BECs with uniform superfluid phase. In our sample preparation, condensates contain fewer than one vortex for $t_{q}$ exceeding 8 s, suggesting that such samples can be considered to have a uniform superfluid phase, i.e., their spatial coherence extends over the size of the sample, $\ell \sim 2R_{y} \approx 300\, \mu$m. For such samples with correlation lengths larger than the intrinsic decay length scale $(\ell \gg \ell_{bg})$, the decay of the interferometry signal is largely dominated by $\ell_{bg}$. As a result, to accurately and reliably determine $\ell_{bg}$, we rely on fully coherent samples (with $t_{q} = 12$ s), where the decay of the signal can be fully attributed to $\ell_{bg}$. Fitting Eq.~(\ref{eq:fitfun}) to the interferometry data for $t_{q} = 12$~s, shown in Fig.~\ref{fig:Braggsetup}(e), we obtain $\ell_{\text{bg}} = 8.3(2)~\mu$m. With this value of $\ell_{bg}$ and the associated uncertainty, we determine the correlation lengths $\ell$ from the interferometry data using Eq.~(\ref{eq:fitfun}).

This intrinsic decay of the interferometry signal is due to the expansion dynamics and the associated phase profile of the condensate. 
A BEC trapped in a harmonic potential exhibits the characteristic parabolic density distribution, which, as the condensate is released from the trap and undergoes free expansion, leads to the emergence of a quadratic spatial phase profile~\cite{castin96}. In our experiment, when two spatially separated, expanding BECs overlap, the quadratic phase profile results in the appearance of linear interference fringes in the density distribution of the condensates~\cite{andrews97}. As multiple interference fringes emerge with increasing separation $\Delta r$, the fractional population of the out-coupled condensate quickly approaches 50\%, leading to a significant reduction in the interference signal, as noted in ~\cite{hagley99,trippenbach20}.

To demonstrate this signal decay mechanism, in Figs.~\ref{fig:fringes}(a) and \ref{fig:fringes}(b), we present the experimental images for samples with large separations of $\Delta r = 20$ and $40\,\mu$m, respectively, where the initial BECs were prepared with $t_q = 10$~s. The spatial distributions of the density difference between the out-coupled and the remaining clouds at their corresponding positions, displayed in Figs.~\ref{fig:fringes}(c) and ~\ref{fig:fringes}(d), reveal the presence of spatial interference fringes in the condensates. We also observe that the key features of the interference fringes, such as their direction and spatial frequency, are adequately captured by numerical simulations [Figs.~\ref{fig:fringes}(e) and ~\ref{fig:fringes}(f)] (see Appendix~\ref{appb} for details). The contrast of these spatial fringes could potentially be utilized to estimate the spatial phase coherence of the condensate, which, however, was not pursued in this work. Previous studies on phase fluctuations in elongated BECs trapped in harmonic potentials have reported similar observations~\cite{hugbart05}.

\subsection{Correlation length versus vortex density}

In the presence of quantum vortices, the superfluid phase of a condensate is not spatially uniform, but varies according to the vortex configuration. Therefore, the phase coherence length $\ell$ of a condensate containing vortices should decrease with increasing vortex density $n_v$. Since the condensate wave function changes sign when crossing a quantum vortex core, it is reasonable to expect that $\ell$ should be linearly proportional to the mean distance between the vortices, $d_v=n_{v}^{-1/2}$. In this section, we examine the relationship between $\ell$ and $n_v$.

The measurement of the number of vortices in a quenched condensate is done by taking a time-of-flight image of the condensate, without applying the Bragg laser beams. Two example images of the condensates after quenching with $t_{q} = 0.5$ s and 4~s are shown in Figs.~\ref{fig:corrlen_vs_qtime}(a) and ~\ref{fig:corrlen_vs_qtime}(b), respectively. In these images, the quantum vortices that are created appear as density-depleted holes in the condensates, whose positions are denoted by white dots. We count the vortices using a machine-learning-based vortex counting algorithm developed and optimized for our system~\cite{kim23}. In Fig.~\ref{fig:corrlen_vs_qtime}(e), the mean number of vortices $N_{v}$ (black squares) is displayed as a function of $t_{q}$ for $t_h=1.25$~s. As the quench rate increases with decreasing $t_q$ from 4~s to 0.5~s, $N_v$ increases from about 3 to more than 20. The number of vortices for fast quenches saturates, as observed in previous experiments~\cite{Goo21,rabga23}. The dashed line in Fig.~\ref{fig:corrlen_vs_qtime}(e) represents the power-law fit to the data within the shaded region ($t_q>1.4$~s) of the form $N_{v} = N_{v,0}t_{q}^{-\alpha^{}_{\text{KZ}}}$, with $\alpha^{}_{\text{KZ}} = 1.7(1)$. More details on the fitting procedure, the determination of the shaded region, and the implications of the measured KZ scaling exponent can be found in \cite{rabga23}.

\begin{figure*}[t]
    \centering
    \includegraphics[width=7in]{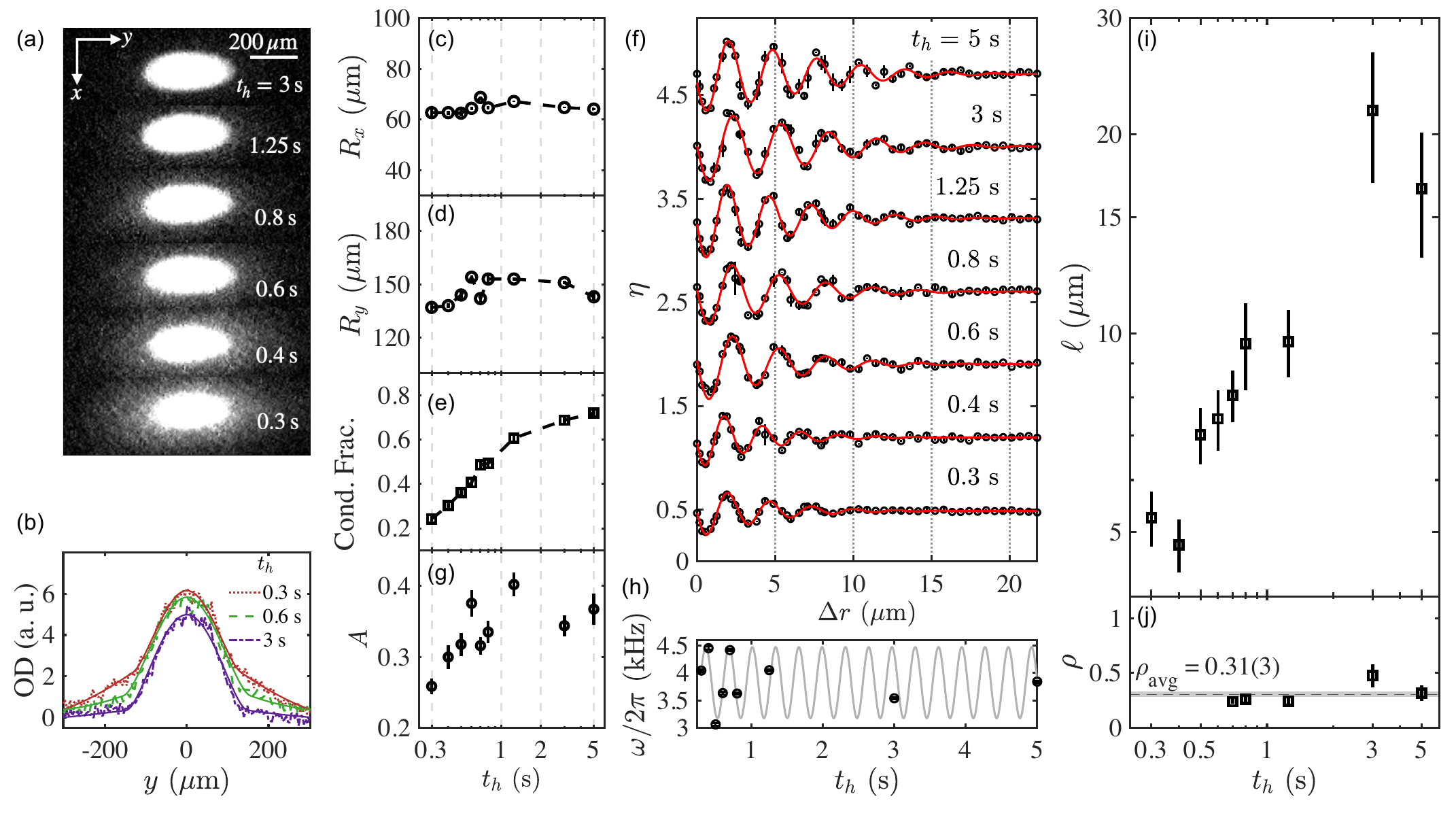}%
    \caption{Early phase coarsening in a rapidly quenched Bose gas. (a) \textit{In situ} images of trapped Bose gas samples rapidly quenched for $t_{q} = 0.5$~s and after variable hold times $t_{h}$. The images are saturated to highlight the thermal component of the samples. (b) Optical density (OD) profiles, integrated along the \textit{x} direction, for various $t_{h}$. The areas under the profiles correspond to the total atom numbers in these samples. As can be seen, the atom number decreases with increasing $t_{h}$. (c) TF radii $R_{x}$ and (d) $R_{y}$ of the condensates, and (e) the condensate fraction as functions of the hold time $t_{h}$. (f) Interferometry signals $\eta$ as functions of $\Delta r$ obtained for samples prepared with different $t_{h}$ values. There is a relative offset of 0.7 between consecutive $t_{h}$ data sets for better visibility. The red line indicates the fit to the data using Eq.~(\ref{eq:fitfun}). (g) Amplitude $A$, and (h) oscillation frequency $\omega$ of the interferometry signal versus $t_{h}$. The gray line in (h) is a sinusoidal curve at a frequency of 3.1 Hz. (i) Coherence length $\ell$ and (j) $\rho=\ell/d_v$ as functions of $t_{h}$. In (j), the dashed line indicates the mean value of $\rho$ and the shaded region indicates the standard error of the mean.}
    \label{fig:corrlen_vs_thold}
\end{figure*}

We measure the coherence lengths $\ell$ for the quenched samples with different $t_q$ using the method described in the previous section. In Figs.~\ref{fig:corrlen_vs_qtime}(c) and ~\ref{fig:corrlen_vs_qtime}(d), we present the interferometry signals $\eta(\Delta r)$ obtained from samples prepared with $t_q=0.5$~s and 4~s, respectively. As expected, the signal decays more rapidly for samples with a higher number of vortices (smaller $t_q$). The measured values of $\ell$ are shown in Fig.~\ref{fig:corrlen_vs_qtime}(e) (red circles). It is clear that the coherence length decreases as the vortex number increases, supporting the notion that a larger defect density leads to a less uniform superfluid phase across the sample, resulting in a smaller spatial coherence length. Furthermore, we observe a plateau in $\ell$ for fast quenches ($t_q < 1$ s), consistent with the saturation of the vortex numbers observed in this regime.

To investigate the dependence of $\ell$ on $d_v$, in Fig.~\ref{fig:corrlen_vs_qtime}(f), we plot the ratio $\rho=\ell/d_v$ using the data obtained from our measurements. The mean intervortex distance is calculated as $d_v=n_v^{-1/2}$ with $n_{v}=N_v/(\pi R_x R_y)$. 
We observe that the ratio remains largely unchanged over the range of $t_q$ values, with a mean value of $\rho_{\text{avg}} = 0.32(4)$, demonstrating the linear relationship between $\ell$ and $d_v$. The value of $\rho_{\text{avg}}$ appears to be reasonable considering the change in sign of the superfluid phase across a vortex core. Based on this finding, we can conclude that the measured values of $\ell$ are reliable indicators of the phase coherence lengths of the quenched condensates, which validates our analysis of the inteferometry signal including the background decay length $\ell_\text{bg}$.

\subsection{Early coarsening effect}

As mentioned earlier, the saturation of vortex number in the fast quench regime represents a departure from the standard KZ description of spontaneous defect formation. This deviation has been attributed to phase coarsening that occurs in the early period of condensate growth prior to vortex formation~\cite{Goo21,Gooetal22,chesler15}. It should be noted that a previous experiment showed that the vortex number statistics is Poissonian even in the saturation regime~\cite{Goo21}. This means, assuming that vortex formation is a random process, that the early phase coarsening is not a result of the annihilation of vortices that are already present in the condensate, but rather an enhanced coarsening process that the incipient, turbulent condensate undergoes before vortices are formed. Therefore, to investigate the underlying defect saturation dynamics, it is important to directly probe the evolution of phase correlations during the early period of phase transition, where a condensate begins to nucleate and grow.

We investigate the early coarsening effect by rapidly quenching samples for $t_{q} = 0.5$~s and changing the post-quench hold time $t_{h}$. As shown in Fig.~\ref{fig:corrlen_vs_qtime}(e), this quench rate is well within the vortex saturation regime~\cite{rabga23}. In Fig.~\ref{fig:corrlen_vs_thold}(a), we present the \textit{in-situ} images of such rapidly quenched samples for different values of $t_{h}$. A condensate grows from the surrounding thermal cloud with increasing $t_{h}$. This is further highlighted in Fig.~\ref{fig:corrlen_vs_thold}(b), where the spatial distributions of the optical density integrated along the $x$ direction show that the thermal fraction of the sample reduces with increasing $t_{h}$. We characterize this condensate growth after rapid quench by measuring the Thomas-Fermi radii $R_{x,y}$ and the condensate fraction using a bimodal fit to the integrated column density profiles. We find that the overall spatial extent of the condensate shows little variation over the range of $t_{h}$ values [Figs.~\ref{fig:corrlen_vs_thold}(c) and \ref{fig:corrlen_vs_thold}(d)] while the condensate fraction increases significantly from $\approx 20\%$ to $70\%$ [Fig.~\ref{fig:corrlen_vs_thold}(e)]. This shows that the condensate grows simultaneously over a large sample area, which is remarkably different from the growth behavior at equilibrium in a trapping potential.

Next, we examine the evolution of the coherence length $\ell$ of the sample as the condensate grows. Figure~\ref{fig:corrlen_vs_thold}(f) shows the inteferometry data $\eta(\Delta r)$ obtained for different values of $t_{h}$. Since the size of the condensate remains relatively constant throughout the range of $t_{h}$ values, despite the increasing condensate fraction, we assume that $\ell_\text{bg}$ does not change significantly with $t_{h}$ and use the same predetermined value as before. We clearly observe that $\ell$ increases as the hold time $t_h$ increases [Fig.~\ref{fig:corrlen_vs_thold}(i)]. Specifically, as $t_h$ is increased from 0.3~s, where stable vortices are yet to form in the sample, to 1.25~s, and as the condensate fraction increases from $\approx 20\%$ to $\approx 70\%$, $\ell$ increases by more than a factor of 2, indicating the coarsening of the phase domains in the sample. This provides a direct evidence for the early coarsening of the order parameter in a rapidly quenched sample, particularly in the absence of stable defect densities, which represents a coarsening mechanism that currently lies beyond the standard Kibble-Zurek descriptions.

We extend the comparison of $\ell$ with the mean intervortex distance $d_v$ to these rapidly quenched samples. As shown in Fig.~\ref{fig:corrlen_vs_thold}(j), for $t_{h}\geq0.7$~s, where vortices could be counted reliably, the length factor $\rho$ shows little variation with $t_{h}$, with an average value of $\rho_{\text{avg}} = 0.31(3)$, which is consistent with the value obtained in Fig.~\ref{fig:corrlen_vs_qtime}(f). 
Utilizing this relationship of $n_v=(\rho/\ell)^2$, we can estimate the number of vortices that may be generated at short hold times when the condensate is too turbulent to have stable vortices. The decrease in coherence length, from a hold time of $t_{h} = 1.25$~s to $t_{h} = 0.3$~s, suggests a corresponding increase in vortex number by a factor of $\ell^{2}_{t_{h} = 1.25\text{s}}/\ell^{2}_{t_{h} = 0.3\text{s}} \approx 3.4$, giving $N_v\approx 70$. Interestingly, this vortex number is close to the number extrapolated from the power-law fit in Fig.~\ref{fig:corrlen_vs_qtime}(e) (indicated by the red square). This observation highlights the presumed role of the early coarsening of the order parameter in defect suppression and saturation in such rapidly quenched samples.

Finally, we discuss the variations in two of the fit parameters of interest, $A$ and $\omega$ with respect to $t_{h}$. We find a strong correlation between the amplitude parameter $A$ and the condensate fraction of the samples [Fig.~\ref{fig:corrlen_vs_thold}(g)]. This indicates that the condensate fraction of the sample determines the amplitude of the interferometry signal. This is consistent with the observations in previous works~\cite{hagley99,Navon15}, where a significant thermal fraction in the sample is expected to result in a rapid initial decay of the signal contrast. Additionally, as shown in Fig.~\ref{fig:corrlen_vs_thold}(h), we observe a temporal modulation of the oscillation frequency $\omega$ with $t_{h}$ at a frequency of 3.1~Hz, which is very close to the trap frequency along the long-axis ($y$) direction, $\omega_{y}/2\pi = 3.2$ Hz. We attribute this modulation of $\omega$ with $t_h$ to the dipole oscillations of the sample caused by the rapid change in the trapping potential during the quench process. The modulation amplitude of $\Delta\omega\approx 2\pi \times 0.7$~kHz corresponds to a velocity of about 2 mm/s for the atoms. This finding exemplifies another advantageous aspect of the off-resonance matter-wave interferometry technique.

\section{Conclusion}

We introduced a method for studying the spatial phase correlations in quenched Bose gases based on off-resonant homodyne matter-wave interferometry. We observe a linear relationship between the measured coherence length $\ell$ and the mean intervortex distance $d_v$, which provides confidence both in our determination of $\ell_{bg}$ and the measured values of $\ell$. In the case of rapidly quenched samples that exhibit vortex number saturation, we investigated the condensate growth dynamics, and provided direct evidence for the early coarsening of the phase domains prior to vortex formation. These findings support the early-coarsening effect, as an explanation for defect saturation, as suggested by previous experiments~\cite{Gooetal22,rabga23}.

This early phase coarsening in an incipient, turbulent condensate, which lies beyond the conventional KZM, is important for a quantitative understanding of the resultant defect density in samples undergoing critical phase transitions. We expect that the matter-wave interferometry method, which is sensitive to decoherence and phase coarsening mechanisms, can be used to study superfluid phase transition dynamics in more detail. We plan to extend this technique to homogeneous samples of Bose gas and make a quantitative comparison of the growth of the coherence length to the resultant vortex numbers, which would provide further insight into the dynamics of spontaneous vortex formation. Although the coherence length and the vortex number were shown to obey the KZ power-law scaling in homogeneous samples in separate systems, a comparative analysis within a single system is still lacking. Moreover, the density homogeneity in such samples mitigates the emergence of a quadratic phase profile over the condensate and the resultant decay of the interference signal, which should greatly simplify the data analysis and allow for a quantitative study of superfluid turbulence in the non-equilibrium phase transition dynamics.

\begin{acknowledgements}
This work was supported by the National Research Foundation of Korea (Grant No. NRF-2023R1A2C3006565) and the Institute for Basic Science in Korea (Grant No. IBS-R009-D1).
\end{acknowledgements}

\appendix

\section{Model function for the interferometry signal}
\label{appa}

Here we describe how the model fit function to the interferometry signal is obtained. Our treatment is similar to what is presented in~\cite{Navon15}. In the matter-wave interferometry using Bragg scattering, we consider the atom-photon interactions between the three bare atomic states $\{\ket{g}, \ket{i}, \ket{e}\}$ and the two light fields characterized by their frequencies $\omega_{1}$ and $\omega_{2}$ [Fig.~\ref{fig:Braggsetup}(a)]. The two light fields drive the $\ket{g} \rightarrow \ket{i}$ and the $\ket{e} \rightarrow \ket{i}$ transitions with Rabi frequencies $\Omega_1$ and $\Omega_2$, respectively, and the two atomic states, $\ket{g}$ and $\ket{e}$ are effectively coupled by the two-photon Bragg scattering process. The energies of the three states are given by $E_{g} = 0$, $E_{i} = \hbar \omega_{i}$, and $E_{e} = \hbar \omega_{e}$ ($0<\omega_{e}\ll \omega_{i}$). $\Delta = \omega_{1}-\omega_{i} (\gg \Omega_{1,2})$ denotes the frequency detuning of the laser beams for the single-photon transitions and $\delta = \omega_{1}-\omega_{2}-\omega_{e}$ denotes the frequency detuning with respect to the resonant two-photon transition.

As shown in  Fig.~\ref{fig:Braggsetup}(c), the matter-wave interferometry consists of three steps that are as follows: splitting by the first Bragg pulse, free propagation for a time $\Delta t$, and recombination by the second Bragg pulse. 
The evolution of the atomic system from the initial state $\ket{\psi_i}$ to the final state $\ket{\psi_f}$ due to the interferometry sequence can be described as 
\begin{equation}
    \ket{\psi_{f}} = \hat{B}(\tau)\hat{U}(\Delta t)\hat{B}(\tau) \ket{\psi_{i}},
\end{equation}
where $\hat{B}(\tau)$ and $\hat{U}(\Delta t)$ are the operators representing the action of a Bragg pulse with duration $\tau$, and the free propagation during the pulse separation time $\Delta t$. 

When the state of the system is represented by $|\psi\rangle=(\psi_g, \psi_e)^\text{T}$, where $\psi_{g(e)}(\vec{r})$ is the wave function of the condensate component in the $|g\rangle$ ($|e\rangle$) state, the Bragg pulse operator, under the rotating wave approximation, is given by
\begin{align}
\begin{split}
    \label{eq:braggoperator}
    \hat{B}(\tau) = \setlength{\arraycolsep}{1pt}
			        \renewcommand{\arraystretch}{1}
           \begin{pmatrix}
\alpha& \beta\\
\beta & \alpha^{*}
\end{pmatrix}
\end{split}
\end{align}
with $\alpha = \cos(\frac{\theta}{2})+i\frac{\delta'}{\Omega_{\mathrm{eff}}} \sin(\frac{\theta}{2})$ and $\beta = -i\frac{\Omega_{\mathrm{R}}}{\Omega_{\mathrm{eff}}} \sin(\frac{\theta}{2})$, where $\theta = \Omega_{\mathrm{eff}}\tau$,  $\Omega_{\mathrm{eff}} = \sqrt{\Omega_{\mathrm{R}}^{2}+\delta'^{2}}$, $\Omega_{\mathrm{R}} = \frac{\Omega_{1}\Omega_{2}}{2\Delta}$, and $\delta' = \delta - \frac{\Omega^{2}_{2}}{4\Delta} + \frac{\Omega^{2}_{1}}{4\Delta}$. 
The free propagation operator is given by 
\begin{equation}
\label{eq:Aoperator}
    \hat{U} (\Delta t)= \left(\begin{matrix}
e^{i\frac{\delta}{2}\Delta t} & 0\\
0 & e^{-i\frac{\delta}{2}\Delta t}\end{matrix}\right)\left(\begin{matrix}
1 & 0\\
0 & \hat{T}_{\Delta t}\end{matrix}\right),
\end{equation}
where $\hat{T}_{\Delta t}$ is the spatial translation operator which acts on the wave function as: $\hat{T}_{\Delta t}\psi(\vec{r},t) = \psi(\vec{r}-\Delta \vec{r}, t)$, where $\Delta \vec{r} = \vec{v}_{r}\Delta t$, and $\vec{v}_{r}$ is the recoil velocity. 

In the final state, the number density of the atoms in the state $\ket{e}$ is given by $n_{r}(\vec{r}) = \abs{\bra{e}\ket{\psi_f}}^{2}$. For the initial state 
$\ket{\psi_{i}} = (\psi_{0}(\vec{r}), 0)^\text{T}$, it is expressed as
\begin{align}
\begin{split}
\label{eq:nv2}
    n_{r} (\vec{r}) =& \abs{\alpha \beta}^{2} \left\{\abs{\psi_{0}(\vec{r})}^{2} + \abs{\psi_{0}(\vec{r}-\Delta \vec{r})}^{2} + \right.\\
    &\left. + 2 \text{Re}\big[ e^{-i(\delta \Delta t+\phi')} \psi_{0}^{*}(\vec{r})\psi_{0}(\vec{r}-\Delta \vec{r})\big] \right\},
\end{split}
\end{align}
where we use $\alpha = \abs{\alpha}e^{i\phi'}$.
Then, the total number of atoms in the out-coupled condensate is given by
\begin{align}
\begin{split}
\label{eq:Nv}
    N_{r} &= \int \text{d}^{3}r \, n_{r} \\
    &= 2\abs{\alpha \beta}^{2} N \left\{1 + C_1(\Delta\vec{r}) \cos{\left(\delta \Delta t+\phi\right)} \right\}
\end{split}
\end{align}
with 
\begin{align}
C_1(\Delta\vec{r})= \frac{1}{N}\abs{\int \text{d}^{3}r \,\psi_{0}^{*}(\vec{r})\psi_{0}(\vec{r}-\Delta \vec{r})},
\end{align}
where $N=\int \text{d}^{3}r \abs{\psi_{0}(\vec{r})}^{2}$ is the total atom number of the condensate. 
$C_1(\Delta\vec{r})$ is the first-order two-point correlation function, which is the quantity of interest, and, in general, is characterized by the correlation length $\ell$ of the system as $C_{1}(\Delta \vec{r}) = \exp(-\Delta r/\ell)$~\cite{Navon15}.

In a previous work ~\cite{hagley99,trippenbach20}, the measured correlation function $C_{1}$ for a BEC trapped in a harmonic potential was observed to be modulated by an envelope function, shown to have a Gaussian form. This envelope function defines a characteristic phase coherence length scale $\ell_{c}$ for the sample. For condensates held in a trap, it represents the decay of the signal with decreasing spatial overlap between the two condensate copies. Consequently, the measured $\ell_{c}$ is observed to be about the characteristic size of the condensate $R_{\text{TF}}$, indicating a BEC with a spatially uniform phase. However, for freely expanding condensates, with substantial spatial phase variations, $\ell_{c} \ll R_{\text{TF}}$, indicating the drastically reduced temporal and spatial phase coherence in the BEC~\cite{trippenbach20}.

From Eq.~(\ref{eq:Nv}) and with the above considerations, we model a fitting function of the interferometry signal $\eta=N_r/N$ as
\begin{equation}
\label{eq:DFfinal}
    \eta(\Delta r) = A \exp{-\left(\frac{1}{2}\frac{\Delta r^{2}}{\ell_{\text{bg}}^{2}} + \frac{\Delta r}{\ell}\right)}\cos\left(\frac{\omega}{v_{r}}\Delta r + \phi\right) + B,
\end{equation}
where, $\ell_{\text{bg}}$ is the characteristic length scale that describes the decay of $\eta$ due to the changing spatial overlap of the condensates and the spatial phase variations that develop in an expanding condensate (Appendix~\ref{appb}). Additionally, as indicated in \cite{hagley99} and as observed in \cite{Navon15}, any non-zero thermal fraction in the sample results in a rapid initial decay of the signal contrast. As a result, we fit for the amplitude $A$ and the signal offset $B$ independently. This is also borne out by the observed dependence of $A$ on the sample condensate fraction, as shown in Fig.~\ref{fig:corrlen_vs_thold}(g). 

\section{Interference fringes in BECs}
\label{appb}

A BEC released from a harmonic trap at $t=0$, develops parabolic spatial phase modulations $\Phi(\vec{r},t) = \sum_{i}c_{i}(t)r_{i}^{2}$ ($i=x,y,z$)~\cite{hugbart05}, and the coefficients $c_{i}(t)$ are given by 
\begin{equation}
    c_{i}(t) = \frac{m}{2\hbar}\frac{\dot{\lambda}_{i}(t)}{\lambda_{i}(t)},
\end{equation}
where $\lambda_{i}(t)$ is the dynamic scaling factor that describes the expansion of the condensate along $\hat{r}_{i}$ direction~\cite{castin96}. In the case where the condensate expands much faster along the $z$ direction compared to the other two directions ($\omega_{z}\gg \omega_{x,y}$), the scaling factors along $k \in (x,y)$ is well approximated by~\cite{castin96}
\begin{equation}
    \lambda_{k}(\tau) = 1+\epsilon_{k}^{2}[\tau \arctan\tau - \ln{\sqrt{1+\tau^{2}}}]+O(\epsilon_{k}^{4}), 
\end{equation}
where $\tau = \omega_{z}t$, and $\epsilon_{k} = \omega_{k}/\omega_{z}$.
Keeping terms upto $O(\epsilon^{2})$, we find that 
\begin{equation}
\label{eq:phasecoeffs}
    c_{k}(t) = \left(\frac{m\omega_{z}}{2\hbar}\right)\epsilon_{k}^{2}\arctan(\omega_{z}t).
\end{equation}

For $\psi_0(\vec{r})=\sqrt{n(\vec{r})}e^{i \Phi(\vec{r},t)}$ and using Eq.~(\ref{eq:nv2}) with $|\alpha\beta|=\frac{1}{2}$, the density distribution of the out-coupled copy is calculated as
\begin{equation}
\label{eq:simdensity}
    n_{r}(\vec{r}) = \frac{n_{-}}{4}+\frac{n_{+}}{4}+\frac{\sqrt{n_{-}n_{+}}}{2}\cos[\delta\Delta t + \Delta \Phi(\vec{r},t)+\phi],
\end{equation}
where $n_{-} = n(\vec{r})$, $n_{+} = n(\vec{r}-\Delta \vec{r})$, and $\Delta \Phi(\vec{r},t) = [c_{x}(t)x\cos(\frac{\pi}{8}) + c_{y}(t)y\sin(\frac{\pi}{8})]\Delta r$ is the phase difference between the two copies. Using Eqs.~(\ref{eq:phasecoeffs}) and (\ref{eq:simdensity}), and the estimated $n(\vec{r})$ from our experimental condition, the density distributions $n_r(\vec{r})$ are calculated for two different sample separations of $\Delta r = 20$ and $40\,\mu$m, and shown in Fig.~\ref{fig:fringes}(c) and \ref{fig:fringes}(d), respectively.

\end{document}